\documentclass[longbibliography,twocolumn,amsmath,amssymb,amsfonts,superscriptaddress,floatfix,showpacs]{revtex4-1}
\usepackage[utf8]{inputenc}
\usepackage[english]{babel}
\usepackage[colorlinks=true,citecolor=green,linkcolor=red,urlcolor=blue]{hyperref}
\usepackage{braket}
\usepackage{bbm}
\usepackage{graphicx}
\usepackage{xcolor}

\newcommand{\F}[1]{{}^#1\!F}
\newcommand{\Fi}[1]{{}_#1\mspace{-1mu}F}

\newcommand{\rh}[1]{{}_#1\mspace{-1mu}\rho}

\newcommand{\skippart}[1]{}
\newcommand{\simplesection}[1]{\emph{#1} ---}

\begin{document}

\title{A critical lattice model for a Haagerup conformal field theory}
\author{Robijn~\surname{Vanhove}}
\affiliation{Department of Physics and Astronomy, Ghent University, Krijgslaan 281, S9, B-9000 Ghent, Belgium}
\author{Laurens~\surname{Lootens}}
\affiliation{Department of Physics and Astronomy, Ghent University, Krijgslaan 281, S9, B-9000 Ghent, Belgium}
\author{Maarten~\surname{Van Damme}}
\affiliation{Department of Physics and Astronomy, Ghent University, Krijgslaan 281, S9, B-9000 Ghent, Belgium}
\author{Ramona~\surname{Wolf}}
\affiliation{Institute for Theoretical Physics, ETH Z\"urich, Wolfgang-Pauli-Str.\ 27, 8093 Z\"urich, Switzerland}
\affiliation{Institut f\"ur Theoretische Physik, Leibniz Universit\"at Hannover, Appelstr.\ 2, 30167 Hannover, Germany}
\author{Tobias J.~\surname{Osborne}}
\affiliation{Institut f\"ur Theoretische Physik, Leibniz Universit\"at Hannover, Appelstr.\ 2, 30167 Hannover, Germany}
\author{Jutho~\surname{Haegeman}}
\affiliation{Department of Physics and Astronomy, Ghent University, Krijgslaan 281, S9, B-9000 Ghent, Belgium}
\author{Frank~\surname{Verstraete}}
\affiliation{Department of Physics and Astronomy, Ghent University, Krijgslaan 281, S9, B-9000 Ghent, Belgium}

\begin{abstract}

We use the formalism of strange correlators to construct a critical classical lattice model in two dimensions with the \emph{Haagerup fusion category} $\mathcal{H}_3$ as input data. We present compelling numerical evidence in the form of finite entanglement scaling to support a Haagerup conformal field theory (CFT) with central charge $c=2$. Generalized twisted CFT spectra are numerically obtained through exact diagonalization of the transfer matrix and the conformal towers are separated in the spectra through their identification with the topological sectors. It is further argued that our model can be obtained through an orbifold procedure from a larger lattice model with input $Z(\mathcal{H}_3)$, which is the simplest modular tensor category that does not admit an algebraic construction. This provides a counterexample for the conjecture that all rational CFT can be constructed from standard methods.
\end{abstract}

\maketitle

\simplesection{Introduction}
Conformal field theory (CFT) plays a central role throughout the natural sciences, from string theory, to the standard model of fundamental physics, through to the effective description of many body systems at criticality \cite{cardyScalingRenormalizationStatistical1996,francescoConformalFieldTheory1997}. As a consequence, the study of CFTs has been an extremely active and vibrant research area since their introduction \cite{belavinInfiniteConformalSymmetry1984}. Fortunately, the rich symmetries exhibited by CFTs, particularly in $1+1$ dimensions, have enabled dramatic progress in the study of their properties and their classification. A prominent role here is played by rational CFTs (RCFT), which supply a rich family of atomic building blocks for general CFTs. These models are highly constrained and, since the inception of CFT, there has been optimism for their classification \cite{wittenSearchHigherSymmetry1989,mooreLecturesRCFT1990}. Considerable progress toward this goal has been achieved. More specifically, it was shown that the underlying mathematical framework of 2d RCFTs is a modular tensor category (MTC) \cite{moore1989classical} by establishing a holographic map between 3d topological field theory (TFT) and 2d CFT. Conversely, it is unknown whether one can construct a CFT based on any modular tensor category and moreover, if standard CFT constructions, such as orbifolds, cosets and simple-current extensions, can produce all RCFTs when applied to the catalogue of basic rational theories \cite{mooreLecturesRCFT1990,wittenSearchHigherSymmetry1989,evansExoticnessRealisabilityTwisted2011, bischoff2016remark}.

The conjecture that all RCFTs can be produced via standard constructions has always been perhaps too bold as there are a variety of potential counterexamples. One particularly exotic candidate which has risen to recent prominence is a putative (R)CFT whose chiral modular data would be realised by the quantum double $\mathcal{D}\mathcal{H}_3$ (or the Drinfeld center $Z(\mathcal{H}_3)$) of the \emph{Haagerup fusion category} $\mathcal{H}_3$ \cite{evansExoticnessRealisabilityTwisted2011,GrossmanSnyder2012HaagerupCats}. This fusion category arose in the mathematical theory of subfactors \cite{jonesIntroductionSubfactors1997} and has, so far, only been constructed via baroque combinatorial methods \cite{asaedaExoticSubfactorsFinite1999,izumiStructureSectorsAssociated2001,petersPlanarAlgebraConstruction2010}. Obviously, for a Haagerup (R)CFT to manifest as an actual counterexample, one would actually need to construct it, which in itself has proven a unique and fascinating challenge.

The first serious efforts to produce a Haagerup (R)CFT commenced with the work of Jones, who exploited ideas from tensor networks to directly build CFT-like continuum theories from fusion category data \cite{jonesScaleInvariantTransfer2017,jonesUnitaryRepresentationsThompson2014,jonesNogoTheoremContinuum2016}. While this initial idea was ultimately unsuccessful \footnote{It was quickly discovered that resulting continuum theory built on $\mathcal{H}_3$ is trivial \cite{jonesNogoTheoremContinuum2016,klieschContinuumLimitsHomogeneous2018} (and also private communication with Cain Edie-Michell)}, it did open the door to the application of a host of recently developed methods \cite{aasen2016topological, vanhove2018mapping,aasen2020topological,lootens2021matrix} to the quest for a Haagerup CFT. These methods require specific lattice realisations built from the input fusion category data. In particular, considerable effort has been expended in using variational matrix product state (MPS) methods \cite{schollwock2011density,haegeman2017diagonalizing,vanderstraeten2019tangent} applied to anyonic spin chain models --- generalizing the well-known golden chain model \cite{feiguin2007interacting,trebstShortIntroductionFibonacci2008} --- built from the $\mathcal{H}_3$ fusion category data, in particular the $F$-symbols \cite{osborne2019f, huang2020f, JacobGit}. A first extensive exploration of the parameter space of the corresponding anyon chains has so far only produced gapped models \cite{wolfMicroscopicModelsFusion2020}, thus excluding a potential Haagerup (R)CFT \footnote{This has recently changed \cite{huang2021Numerical}}.

The negative results discovered ad hoc in the study of anyon chains motivated the application of the Haagerup case to a larger research programme targeting the systematic construction of full CFTs from topological modular data. This programme is built on the premise that an arbitrary CFT may be microscopically realised via the so-called strange correlator \cite{you2014wave, vanhove2018mapping,vanhoveTopologicalAspectsCritical2021} applied to different tensor network representations \cite{lootens2021matrix} of Levin-Wen string-net models \cite{levinStringnetCondensationPhysical2005,hahnGeneralizedStringnetsUnitary2020,huangTopologicalFieldTheory2021}. This idea is attractive for a variety of reasons as: (1) it provides a systematic way to build the geometric correlation data of a CFT from the purely topological modular data; (2) it supplies a clear and direct realisation of the symmetries of the CFT via matrix product operators (MPO); (3) it enables the direct application of tensor-network methods to study the resulting critical lattice model and enables the selection of topological sectors in a systematic way. The programme is essentially a lattice implementation of the description of the topological aspects (such as partition functions and topological defects) of 2d RCFTs in terms of MTCs and their representations, as described in a series of detailed papers by Fr\"ohlich, Fuchs, Runkel and Schweigert \cite{felder2000conformal, fuchs2002tft, frohlich2004kramers}. The Haagerup (R)CFT presents a fascinating foil for cutting-edge research in tensor-network methods as the underlying fusion category is not braided and the center $Z(\mathcal{H}_3)$ exhibits multiplicities in the fusion spaces \cite{Hong:2007ty}, obstructing the key assumptions of a host of well-developed tools.


In this Letter we report on a critical classical lattice model, obtained via the strange-correlator construction applied to a TFT tensor network with the $\mathcal{H}_3$ fusion category data as input. We argue that this statistical mechanics model is associated to a CFT with modular data $Z(\mathcal{H}_3)$ and a central charge $c=2$, indicating that it is not the conjectured Haagerup (R)CFT in itself (which is understood to have central charge divisible by $8$). We apply a complementary portfolio of numerical methods, including, anyonic infinite variational tensor network methods and announce a general numerical method for selecting topological sectors in critical lattice models with input data of potentially non-braided fusion categories. This method encompasses the special case of modular input categories \cite{Pfeifer:2010xi}. \\

\begin{figure}[hbt!]
\centering
\includegraphics[width=0.8\linewidth,keepaspectratio=true,page=4]{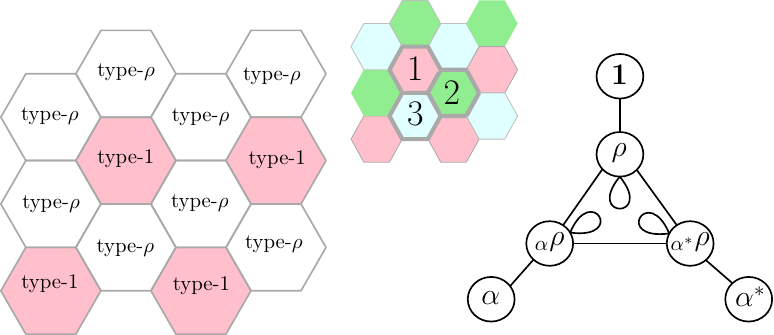} 
\caption{The partition function of the model on the hexagonal lattice in a maximally occupied configuration, i.e. as many type-$\bf{1}$ plaquettes as allowed by the Fibonacci grading of the fusion rules. There are three maximally occupied configurations as indicated by the three different sublattices. The adjacency rules of neighbouring particles can be shown by the corresponding Dynkin diagram.}
\label{fig:hexLattice}
\end{figure}

\simplesection{A strange correlator for the $\mathcal{H}_3$ fusion category}
We construct a two dimensional lattice model starting from generalized projected entangled pair state (PEPS) representations of string-net ground states as described in \cite{lootens2021matrix}. The construction requires two fusion categories $\mathcal{C}$ and $\mathcal{D}$ and a $(\mathcal{C},\mathcal{D})$-bimodule category $\mathcal{M}$. In the remainder we will use the convention that $\mathcal{C}$ labels the symmetries, $\mathcal{D}$ is the input of the string-net construction and $\mathcal{M}$ labels the virtual degrees of freedom of the tensor network. We can build the generalized string-net ground state on a honeycomb lattice using the following trivalent PEPS tensors:   
\begin{align}
\begin{split}
\vcenter{\hbox{\includegraphics[page=1, scale=0.7]{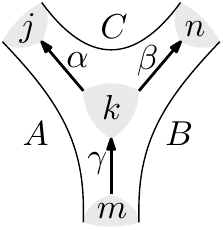}}} := \resizebox{0.22\hsize}{!}{$\frac{ \left(\F{3}^{A\alpha\beta}_B\right)^{\gamma,km}_{C,jn}}{\sqrt{d_C}}$},
\vcenter{\hbox{\includegraphics[page=2, scale=0.7]{figures}}} := \resizebox{0.22\hsize}{!}{$\frac{ \left(\Fi{3}^{A\alpha\beta}_B\right)^{\gamma,km}_{C,jn}}{\sqrt{d_C}}$},
\end{split}
\label{peps}
\end{align}
where $\F{3}$ ($\Fi{3}$) is the (inverse) module associator of $\mathcal{M}$ as a right $\mathcal{D}$ module category with $\{A,B,C\} \in \mathcal{M}$ and $\{\alpha,\beta,\gamma\} \in \mathcal{D}$. For our model we will make the choice $\mathcal{C} = \mathcal{M} = \mathcal{D} = \mathcal{H}_3$ with $\mathcal{H}_3$ the $G=\mathbb{Z}_3$ Haagerup-Izumi category \cite{evansExoticnessRealisabilityTwisted2011, huang2020f}. This choice coincides with the original PEPS representation for string-net ground states \cite{buerschaper2009explicit, bultinck2017anyons}, in which case $\F{3}=F$ is a unitary solution of the pentagon equation and the fusion multiplicities are trivial \cite{osborne2019f, huang2020f, JacobGit}. $\mathcal{H}_3$ has six simple objects $\{\bf{1},\alpha,\alpha^*\}$ (we will call type-$\bf{1}$) and $\{\rho,\rh{\alpha},\rh{{\alpha^{*}}}\}$ (we will call type-$\rho$), with non-trivial fusion rules $\alpha^3 = \bf{1} $, 
\begin{align}
\begin{split}
    \rh{\alpha} = \alpha \otimes \rho, \quad \rh{{\alpha^{*}}} = \alpha^{*} \otimes \rho \\
    \rho \otimes \alpha = \rh{{\alpha^{*}}}, \quad  \rho \otimes \alpha^{*} = \rh{\alpha}
\end{split}
\label{fusion}
\end{align}
and $\rho \otimes \rho = \bf{1} \oplus \rho \oplus \rh{\alpha} \oplus \rh{{\alpha^{*}}}$. The other fusion rules of the form $\text{type-}\rho \otimes \text{type-}\rho$ can be obtained from Eq.~\ref{fusion}.
The fusion rules admit a Fibonacci grading between the type-$\bf{1}$ and the type-$\rho$ objects: $\text{type-}\rho \otimes \text{type-}\rho = \{\text{type-}\bf{1}\} \oplus \{\text{type-}\rho\}$. The corresponding quantum dimensions are $d_{\text{type-}\bf{1}} = 1$ and $d_{\text{type-}\rho} = \frac{3+\sqrt{13}}{2}$. The PEPS tensors have virtual MPO symmetries \cite{bultinck2017anyons}, i.e.\ string-like operators that can be freely pulled through the lattice without any action on the physical indices:
\begin{align}
\begin{split}
\vcenter{\hbox{\includegraphics[page=10, scale=1]{figures}}}.
\end{split}
\label{pullingThrough}
\end{align}
For diagrammatic convenience, we have omitted the triple line notation and will keep doing so going forward, leaving out the loops (capital roman letters) in our diagrams. The MPOs are labeled by small roman letters ($a \in \mathcal{C}=\mathcal{H}_3$).
In the next step, we choose a strange correlator \cite{vanhove2018mapping} by fixing the physical indices (greek letters) of the ground state to $\rho$, obtaining a lattice partition function where the degrees of freedom are the loops of the original PEPS. The resulting partition function will inherit the virtual MPO symmetries of the PEPS, which become the lattice manifestation of the continuum CFT topological defects.
The situation is very similar as in the case of the hard hexagon model \cite{baxter1980hard}, which can be constructed from the Fibonacci string-net ground state (with simple objects $\{\bf{1},\tau\}$) by fixing the physical indices on the non-trivial object $\tau$ \cite{vanhove2018mapping} (the Fibonacci case can be viewed as the $G = \mathbb{Z}_1$ Haagerup-Izumi category). The result of this choice is a hard constraint between neighbouring plaquettes (sites on the dual triangular lattice), not allowing any adjacent plaquettes labeled by the object $\bf{1}$. Similarly for the $\mathcal{H}_3$ model, adjacent plaquettes cannot by occupied by type-$\bf{1}$ objects. The adjacency rules for neighbouring plaquettes can be diagramatically shown as a Dynkin diagram (Fig.~\ref{fig:hexLattice}).

Besides the internal $\mathbb{Z}_3$-symmetry generated by $\alpha$, the model has an extra $\mathbb{Z}_3$ sublattice symmetry. This can be seen in the maximally occupied configuration (the configuration with as many type-$\bf{1}$ objects on the plaquettes as allowed by the hard constraint) (see Fig.~\ref{fig:hexLattice}) \cite{baxter1980hard}. As a consequence of the sublattice symmetry, we have to define the transfer matrix of the model on a ring of length $L=3n, n\in\mathbb{Z}$. Choosing $L\neq3n$ amounts to introducing a non-trivial twist. The $\mathbb{Z}_3$ sublattice symmetry is generated by shifting the lattice by one site and becomes an invertible topological defect line of the CFT in the continuum limit. This is in analogy to the critical hard hexagon case, where the continuum limit is described by the $\mathbb{Z}_3$ parafermion CFT with central charge $c = 4/5$ which indeed has a $\mathbb{Z}_3$ symmetry generated by the one-site shift on the lattice. Similarly, a critical theory can be constructed for the second fusion category in the series (the $G = \mathbb{Z}_2$ Haagerup-Izumi category), the simple objects of which coincide with the integer representations of $su(2)_6$. The continuum theory in this case is described by the $\mathbb{Z}_6$ parafermion CFT with $c=5/4$. Again, the $\mathbb{Z}_3$ subgroup of $\mathbb{Z}_6$ can be generated on the lattice by a one-site shift. The identification of the critical models for the Haagerup-Izumi series to pure $\mathbb{Z}_n$ parafermion CFTs breaks down for $\mathcal{H}_3$.

The topological sectors of the model are given by the Drinfeld center $Z(\mathcal{H}_3)$, which is not simply $\mathcal{H}_3 \boxtimes \mathcal{H}_3^\text{op}$ in this case since $\mathcal{H}_3$ is not even braided, let alone modular. The center has twelve simple objects, labeled by $(Z,\Omega)$, where $Z$ is an object in $\mathcal{H}_3$ and $\Omega$ the half-braiding. If we write $Z$ in the basis of simple objects, we can label the twelve objects as in Table~\ref{centerLabeling}, using a similar labeling as in \cite{evansExoticnessRealisabilityTwisted2011, Hong:2007ty}.
\begin{table}[hbt!]
\begin{tabular}{|c|c|c|c|}
	\hline
	$Z(\mathcal{H}_3)$ & $Z \in \mathcal{H}_3$ & $\Omega$ & dim\\
	\hline
	id&\bf{1}&id$^{1}$&$1$\\
	$\mu^{1,2,3,4,5,6}$&$\rho\oplus\rh{{\alpha}}\oplus\rh{{\alpha^{*}}}$&$\mu^{1,2,3,4,5,6}$&$3d_{\rho}$\\
	$\pi_1$&$\bf{1}\oplus\rho\oplus\rh{{\alpha}}\oplus\rh{{\alpha^{*}}}$&$\pi_1^1$&$3d_{\rho}+1$\\
	$\pi_2$&$\bf{1}\oplus\bf{1}\oplus\rho\oplus\rh{{\alpha}}\oplus\rh{{\alpha^{*}}}$&$\pi_2^1$&$3d_{\rho}+2$\\
	$\sigma^{1,2,3}$&$\alpha\oplus\alpha^{*}\oplus\rho\oplus\rh{{\alpha}}\oplus\rh{{\alpha^{*}}}$&$\sigma^{1,2,3}$&$3d_{\rho}+2$\\
	\hline
\end{tabular}
\caption{The twelve simple objects of $Z(\mathcal{H}_3)$ labeled by an object $Z \in \mathcal{H}_3$ and a half-braiding $\Omega$. The last column is the corresponding quantum dimension. For example, there are six different objects $\mu$ labeled by the same object $Z$ and different half-braidings ($\mu^{1,2,3,4,5,6}$).}
\label{centerLabeling}
\end{table}

The half-braidings of objects in $Z(\mathcal{H}_3)$ are natural, meaning that in particular they satisfy
\begin{equation}
   \vcenter{\hbox{\includegraphics[page=16, scale=0.6]{figures}}},
   \label{naturality}
\end{equation}
with $Z, a,b,c \in \mathcal{H}_3$ and $\Omega$ the specified half-braiding. This property is reminiscent of the pulling-through equation \ref{pullingThrough}, and in fact, when written out using the explicit definitions for these tensors this becomes the hexagon equation. These half-braidings can be obtained from the tube algebra idempotents \cite{lootens2021topological} that project on a topological sector in $Z(\mathcal{H}_3)$, as described above.\\

\begin{figure}[hbt!]
\centering
\includegraphics[width=0.9\linewidth,keepaspectratio=true]{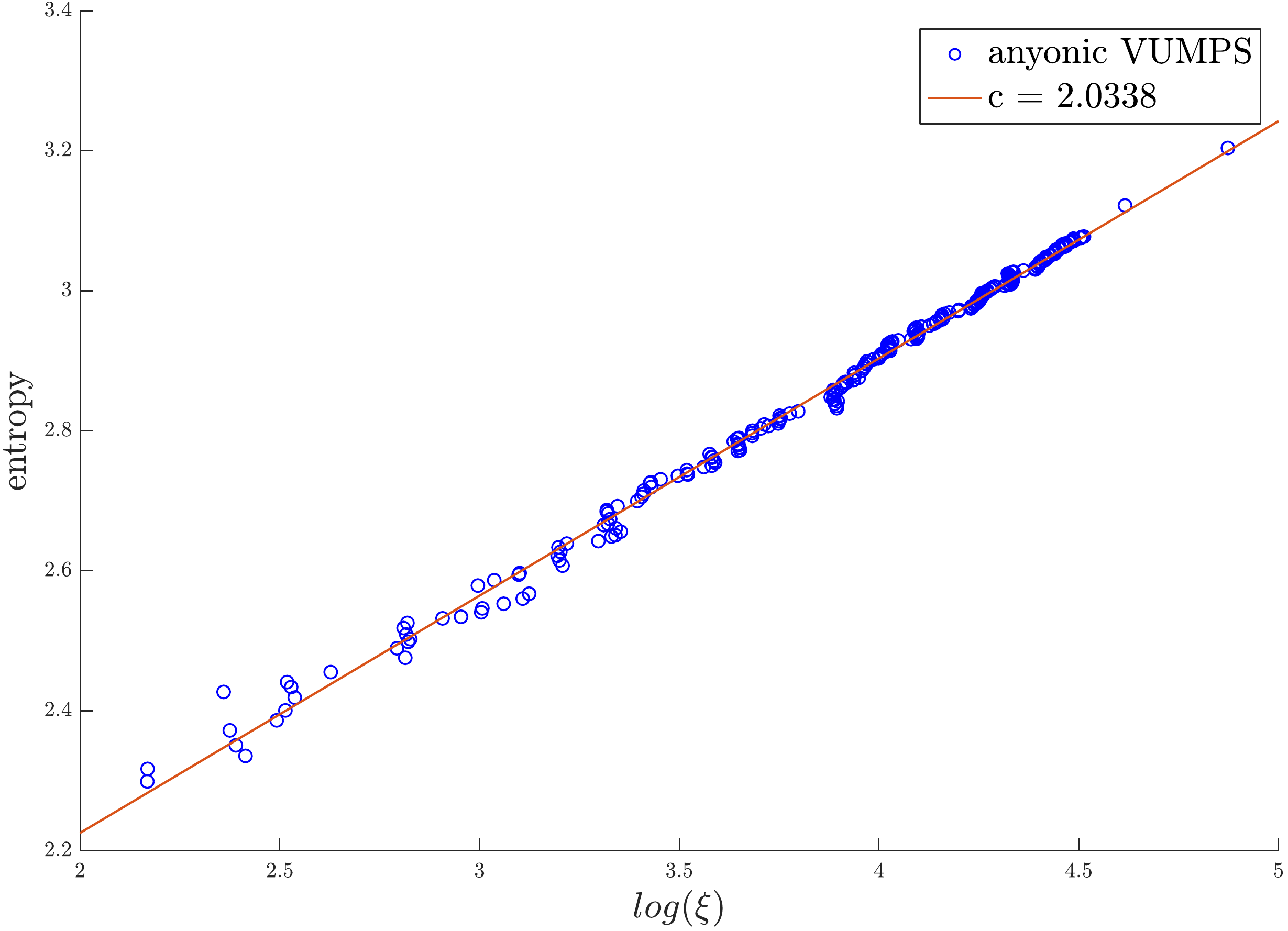} 
\caption{Finite entanglement scaling for the fixed point MPS of the transfer matrix calculated using VUMPS with explicit $\mathcal{H}_3$ anyonic symmetry. The results are consistent with a central charge close to $c=2$.}
\label{scaling}
\end{figure}

\simplesection{Numerical results} 
We have performed variational uniform matrix product state (VUMPS) simulations \cite{zauner2018variational} for the MPS fixed point of the transfer matrix in the thermodynamic limit for increasing bond dimension. The algorithm explicitly preserves the anyonic $\mathcal{H}_3$ symmetry, allowing for higher bond dimensions than standard methods \cite{pfeifer2012simulation, Pfeifer:2010xi, tensorKit, MPSKitMaarten}. As the entanglement entropy in an infinite chain diverges with increasing MPS bond dimension at criticality, it scales as $S = \frac{c}{6}log(\xi)$ \cite{calabrese2006entanglement,tagliacozzo2008scaling}, with $\xi$ the MPS correlation length. The result is shown in Fig. \ref{scaling} up to $\xi \approx 150$ and strongly indicates a critical theory with central charge $c=2$.

\begin{figure}[hbt!]
\centering
\includegraphics[width=0.8\linewidth,keepaspectratio=true,page=17]{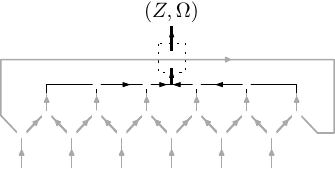} 
\caption{The exact diagonalization scheme with anyonic symmetries (shown for $L=6$). The grey lines of the one-row transfer matrix indicate that they have been fixed on the strange correlator $\rho$. The eigenvector is written as an anyonic fusion tree (black lines). Whenever a line crosses the sector label (the black line going up) a half-braiding is required. The anyonic symmetry is ensured by Eq.~\ref{naturality}.}
\label{ED_halfbraid}
\end{figure}

Secondly, we have performed exact diagonalization with anyonic symmetry \cite{pfeifer2012simulation,tensorKit} on the transfer matrix with periodic boundary conditions. The method consists of writing the eigenvector of the transfer matrix as an anyonic fusion tree, fixing the outgoing legs on the strange correlator $\rho$. The eigenvector is chosen in a specific topological sector $(Z,\Omega)$ by fixing the total charge of the fusion tree to $Z$ and using the half-braiding $\Omega$ whenever a crossing is required (see Fig.~\ref{ED_halfbraid}). The simple objects in the decomposition of $Z = \bigoplus_a a$ for a given sector $(Z,\Omega)$ (see Table \ref{centerLabeling}) indicates the presence of that sector in the spectrum of the transfer matrix twisted by the corresponding topological twists $a$. 

\begin{figure}[hbt!]
\centering
\includegraphics[width=\linewidth,keepaspectratio=true]{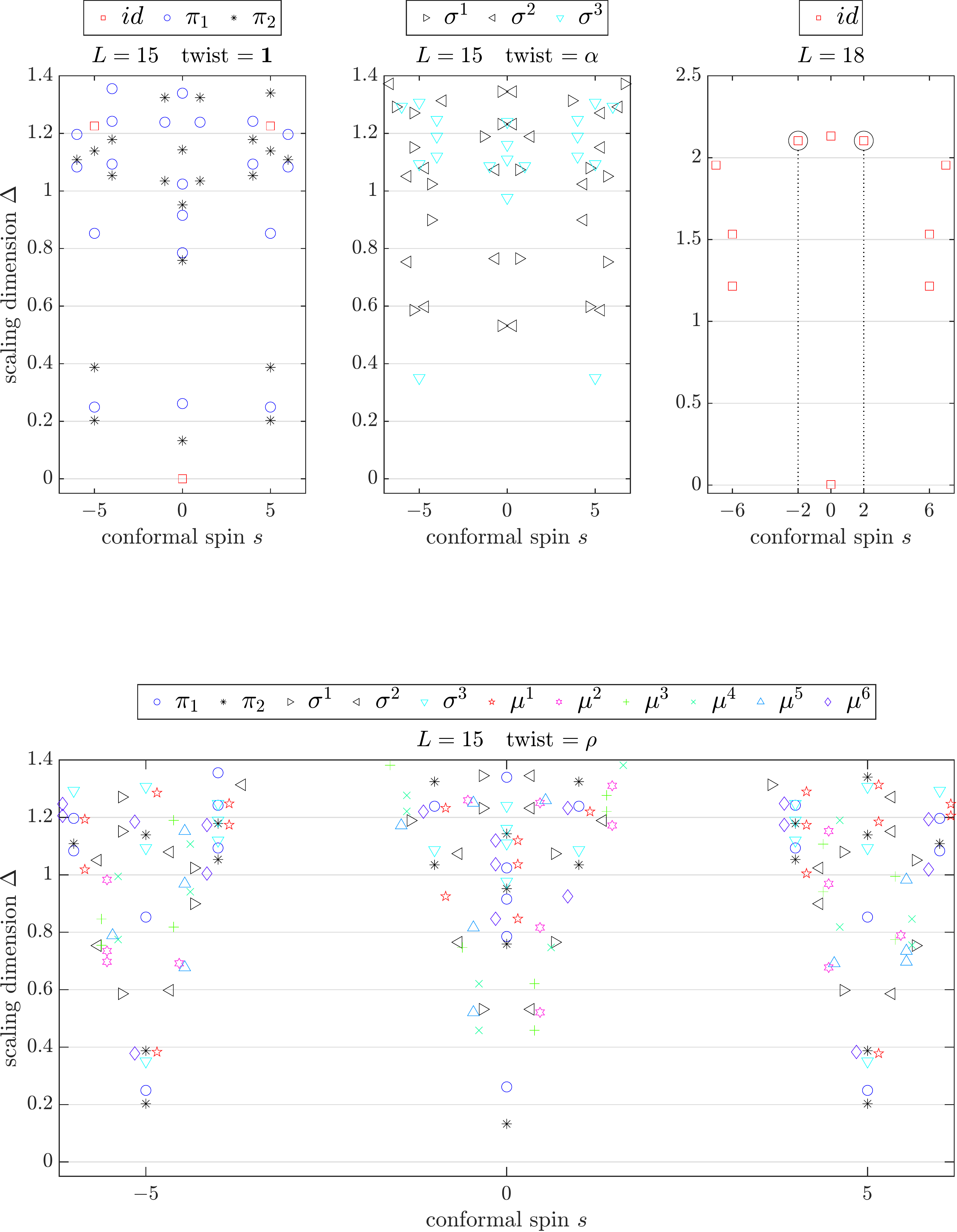} 
\caption{Spectra for the transfer matrix, twisted with topological defects $\bf{1}$ (upper left), $\alpha$ (upper middle) and $\rho$ (bottom), numerically obtained with anyonic symmetry-preserving exact diagonalization on $L=15$ sites. The eigenvalues are labeled by their corresponding topological sectors $Z(\mathcal{H}_3)$ according to Table \ref{centerLabeling}. For the non-trivial twists, the conformal spins are only integers up to a topological spin correction (Table \ref{topoSpins}). Upper right: the identity sector on $L=18$ sites. The first excited states of the vacuum ($\Delta=2, s=-2,2$) are circled in black.}
\label{spectra}
\end{figure}

The spectra of the transfer matrix with a trivial ($a=\bf{1}$), $\alpha$- and $\rho$-twist on $L=15$ sites are shown in Fig.~\ref{spectra}, together with the trivial sector on $L=18$ sites. \footnote{Note that the conformal spins in Fig.~\ref{spectra} are determined under a one-site shift, so they do only correspond to the true conformal spins modulo $L/3$}. The numerically obtained ground state in the trivial sector has a finite-size correction $E_0 \sim fL + \frac{\pi c v}{6L}$ \cite{cardy1986operator}, where $\exp(-f)$ is the free energy per site in the thermodynamic limit and $v$ the characteristic velocity, both of which can be determined by fitting the ground state energy for several sizes $L=6,9,12,15,18$. We label the spectra with the sectors of $Z(\mathcal{H}_3)$ (Table \ref{centerLabeling}). The conformal spins in each sector acquire a topological correction shown in Table \ref{topoSpins} \footnote{The topological spins can be determined by projecting the Dehn twist on a topological sector, see for example \cite{aasen2016topological}}.

\begin{table}[hbt!]
\begin{tabular}{|c||c|c|c|c|c|c|c|c|c|c|c|c|}
	\hline
	$Z(\mathcal{H}_3)$&$id$&$\pi_1$&$\pi_2$&$\sigma^1$&$\sigma^2$&$\sigma^3$&$\mu^1$&$\mu^2$&$\mu^3$&$\mu^4$&$\mu^5$&$\mu^6$\\
	\hline
	\begin{tabular}{@{}c@{}}topological \\ spin\end{tabular}&$0$&$0$&$0$&$-\frac{1}{3}$&$\frac{1}{3}$&$0$&$\frac{2}{13}$&$\frac{6}{13}$&$\frac{5}{13}$&$-\frac{5}{13}$&$-\frac{6}{13}$&$-\frac{2}{13}$\\[1.2ex]
	\hline
\end{tabular}
\caption{Topological spins of the sectors in $Z(\mathcal{H}_3)$.}
\label{topoSpins}
\end{table}

The twisted spectra are expected to show the emerging CFT towers up to finite-size effects. The torus partition function (twisted in one direction by $a$) is of the form $ Z_{a} \simeq  \sum_{\alpha,\bar{\beta}}\chi_{\alpha}(q)\widetilde{M}^{a}_{\alpha\bar{\beta}}\overline{\chi}_{\overline{\beta}}(\overline{q})$ \cite{petkova2001generalised}
and is in particular modular invariant for $a=\bf{1}$. Projecting the spectrum onto a topological sector amounts to breaking down the partition function into single (or possibly sums) of sesquilinear character terms \cite{vanhove2018mapping, vanhoveTopologicalAspectsCritical2021}. The conformal spin $s = h_{\alpha} - h_{\beta}$ ($h$ is the conformal weight) of the lowest lying eigenvalue in the tower $\chi_{\alpha}\overline{\chi}_{\overline{\beta}}$ corresponds to the topological spin of that sector. Note that the multiplicity of the simple object $\bf{1}$ in the sector $\pi_2$ (see Table \ref{centerLabeling}) signals an exact degeneracy in the $\bf{1}-$twisted spectrum and a corresponding multiplicity in a term of the partition function ($2|\chi_{\alpha}|^2$). The partition function is expected to be non-diagonal ($M_{\alpha \beta} \neq \delta_{\alpha,\beta}$), which in particular implies that the first eigenvalue at conformal spin $s=1$ in a certain tower is not necessarily the first excited state of the lowest lying state in that tower, but could in principle be the first state of a tower belonging to a term $\chi_{\alpha}\overline{\chi}_{\overline{\beta}}$ with $h(\alpha)-h(\beta)=1$.

\simplesection{Discussion}
The topological aspects of 2d RCFT can be captured through a holographic principle starting from a 3D TQFT. The Moore-Seiberg data concerning the representations of the chiral algebra of the CFT is described by a MTC $\mathcal{D}$. Therefore, not surprisingly, a critical lattice theory (including the topological defects) can be constructed from a string-net ground state based on a modular tensor category. Constructing a modular invariant can then be done by choosing a right $\mathcal{D}$-module category $\mathcal{M}$. In this way, non-diagonal partition functions ($M_{\alpha \beta} \neq \delta_{\alpha,\beta}$) are constructed by choosing different module categories $\mathcal{M}$. The topological defects are in turn given by objects in a category $\mathcal{C}$, which depends on $\mathcal{D}$ and $\mathcal{M}$ by requiring that $\mathcal{M}$ is an invertible ($\mathcal{C}$,$\mathcal{D}$)-bimodule category. We refer to \cite{lootens2021matrix} for the details regarding the tensor network representations of the corresponding string-net ground states.

In the model described above the choice $\mathcal{D}=\mathcal{M}=\mathcal{H}_3$ was made, but since $\mathcal{H}_3$ is not modular, it does not directly capture the chiral algebra of the underlying CFT. Exactly the same model can be obtained by choosing $\mathcal{D} = Z(\mathcal{H}_3)$ (which is modular) and $\mathcal{M} = \mathcal{H}_3$ by choosing a suitable strange correlator $X$ \footnote{In fact any strange correlator of a model with $\mathcal{D} = \mathcal{M} = \mathcal{A}$, with $\mathcal{A}$ some fusion category, can be written as a strange correlator of a model with $\mathcal{D} = Z(\mathcal{A})$ and $\mathcal{M} = \mathcal{A}$, see \cite{bramNewPaper}}. We have numerically verified this by computing the $\F{3}$ and $\F{4}$-symbols for this bimodule category based on a method described in Sect.~IV of \cite{williamson2017symmetry}. These solutions satisfy the relevant pentagon equations up to machine precision. Due to the current lack of a symbolic expression for these $F$-symbols, we leave their presentation for future work.

Assuming the model we consider should indeed be thought of as a strange correlator $X$ on a PEPS with $\mathcal{D} = Z(\mathcal{H}_3)$ and $\mathcal{M} = \mathcal{H}_3$, the corresponding CFT will not have a diagonal partition function, as this is only the case when $\mathcal{M} = \mathcal{D}$. Indeed, our model (blue) can be locally mapped to one where $\mathcal{M} = \mathcal{D} = Z(\mathcal{H}_3)$ (green) by using an MPO intertwiner $A \in \mathcal{M} = \mathcal{H}_3$:
\begin{equation}
    \vcenter{\hbox{\includegraphics[page=8, scale=1]{figures}}}.
\end{equation}
On the level of the untwisted (empty) torus partition function, an intertwiner loop ($A=\bf{1}$ for example) can be freely grown and wrapped around the torus, mapping the partition function of the original model (blue) to a twisted partition function of the new model (green):
\begin{equation}
    Z = \vcenter{\hbox{\includegraphics[page=13, scale=0.4]{figures}}} = \vcenter{\hbox{\includegraphics[page=14, scale=0.4]{figures}}} = \vcenter{\hbox{\includegraphics[page=15, scale=0.4]{figures}}},
\end{equation}
in which the twist $\alpha$ in the final partition function decomposes as $\alpha = id \oplus \pi_1 \oplus \pi_2 \oplus \pi_2$, implying that our model can be obtained from a diagonal one through a (generalized) orbifold construction \cite{frohlich2010defect}. The enlarged model based on $\mathcal{D} = Z(\mathcal{H}_3)$ shows that the symmetries of the model we study are actually given by $\mathcal{C} = \mathcal{H}_3 \boxtimes \mathcal{H}_3^\text{op}$, implying that we did not consider the full set of topological defects. Doing so requires solving the complete set of pentagon equations for this particular bimodule category, which we plan to explore in future work. This would enable a further decomposition of the spectrum by labeling it with objects in $Z(Z(\mathcal{H}_3)) = Z(\mathcal{H}_3) \boxtimes Z(\mathcal{H}_3)^{\text{op}}$, both for the diagonal and non-diagonal partition function.

We end by noting that the observed central charge of $c = 2$ appears to be in contradiction with an underlying MTC corresponding to $Z(\mathcal{H}_3)$, which should have $c = 0 \mod 8$ \cite{evansExoticnessRealisabilityTwisted2011}. Drawing on experience with other lattice models constructed as a strange correlator however, this situation is not unfamiliar; for example, the critical RSOS models as constructed from the $\text{su}(2)_k$ MTCs (in a certain regime) are not described by a WZW CFT with central charge $3k/(k+2)$, but rather by the minimal models with $c<1$ that can be understood as cosets of $\text{su}(2)_k$ WZW models \cite{Bazhanov:1987zu}. The fact that the coset MTC describing the CFT is not required to construct the critical lattice model can be understood by the fact that the lattice model does not necessarily have all the topological defect symmetries of the continuum CFT, as some of these can be broken by irrelevant perturbations in finite size only to be recovered in the continuum limit \cite{Belletete:2020gst}. This scenario appears to be quite common, and we speculate that the model we study here is no different and that the MTC $\mathcal{D}_\text{coset}$ of the CFT describing our critical lattice model is a coset involving the MTC $Z(\mathcal{H}_3)$ and another MTC with $c = 6$, e.g. $\mathcal{D}_\text{coset} = Z(\mathcal{H}_3)/\mathcal{D}_{c=6}$. The precise nature of this coset requires a detailed analysis of the spectrum, as well as a characterization of the possible cosets involving $Z(\mathcal{H}_3)$.



\simplesection{Conclusion} 
We have shown strong numerical evidence for a Haagerup CFT with central charge $c=2$, using the strange correlator prescription for the Haagerup fusion category $\mathcal{H}_3$. The numerical results were obtained using anyonic symmetry-preserving algorithms with explicit topological sector selection without requiring braiding of the input category. The model admits an interpretation as a non-diagonal modular invariant of a CFT with an MTC corresponding to $Z(\mathcal{H}_3)$. We argue that the observed central charge of the critical lattice model can be obtained as a coset involving $Z(\mathcal{H}_3)$, although an explicit construction requires further analysis. Furthermore, it is worth investigating if similar critical lattice models can be constructed (and their corresponding CFTs identified), for the general series of Haagerup-Izumi fusion categories.


\begin{acknowledgments}

\end{acknowledgments}

Near the completion of this work, we learned that a critical anyonic chain Hamiltonian for the Haagerup fusion category $\mathcal{H}_3$ was obtained by Tzu-Cheng Huang, Ying-Hsuan Lin, Kantaro Ohmori, Yuji Tachikawa and Masaki Tezuka. Numerical evidence also indicates a central charge $c=2$ CFT for this Hamiltonian. A preliminary check indicates that this Hamiltonian is not merely the (1+1)d quantum analogue of the 2D classical model discussed in this work. We thank Tzu-Cheng Huang, Ying-Hsuan Lin, Kantaro Ohmori, Yuji Tachikawa and Masaki Tezuka for insightful discussions and for a coordinated submission of our manuscripts. We are grateful for numerous helpful conversations, and correspondence, with Jacob Bridgeman, Cain Edie-Michell, Paul Fendley, J\"urgen Fuchs, Ash Milsted, Christoph Schweigert, and Sukhwinder Singh. RV and LL are supported by a Fellowship from the Research Foundation Flanders (FWO). RW is supported by the Swiss National Science Foundation via the National Center for Competence in Research \textit{Quantum Science and Technology} (NCCR QSIT) and the NCCR \textit{SwissMAP -- The Mathematics of Physics}. This work has received funding from the European Research Council (ERC) under the European Unions Horizon 2020 research and innovation programme (grant agreements No 647905 (QUTE) and 715861 (ERQUAF)), and from the Research Foundation Flanders via grant GOE1520N. Support by the DFG through SFB 1227 (DQ-mat), Quantum Valley Lower Saxony, and funding by the Deutsche Forschungsgemeinschaft (DFG, German Research Foundation) under Germanys Excellence Strategy EXC-2123 QuantumFrontiers 390837967 is also acknowledged.

\bibliographystyle{apsrev4-1}
\bibliography{haagerup}

\end{document}